\documentclass[a4paper,11pt]{article}
\usepackage[top=0.85in,left=1in,right=1in,footskip=0.75in]{geometry}
\usepackage[T1]{fontenc}
\usepackage{lmodern}

\usepackage{changepage}

\usepackage[utf8x]{inputenc}

\usepackage{fixltx2e}

\usepackage{amsmath,amssymb}

\usepackage{cite}

\usepackage{nameref,hyperref}

\usepackage[right]{lineno}

\usepackage{epsfig}
\usepackage{soul}
\usepackage{xskak}

\newcommand{\avrg}[1]{\left\langle #1 \right\rangle}

\begin{document}

\begin{center}
{\Large
\textbf{A study of memory effects in a chess database} 
}

Ana L. Schaigorodsky\textsuperscript{1,2*},
Juan I. Perotti\textsuperscript{3},
Orlando V. Billoni\textsuperscript{1,2}\\
\end{center}
\vspace{0.2cm}

\textbf{1} Facultad de Matem\'atica, Astronom\'{\i}a,
F\'{\i}sica y Computaci\'on, Universidad Nacional de C\'ordoba, Ciudad Universitaria, C\'ordoba, Argentina.

\textbf{2}  Instituto de F\'{\i}sica Enrique Gaviola (IFEG-CONICET), 
Ciudad Universitaria, C\'ordoba, Argentina.

\textbf{3} IMT School for Advanced Studies Lucca, Piazza San Francesco 19, I-55100, Lucca, Italy.

\vspace{0.5cm}
\begin{flushleft}
* schaigorodsky@famaf.unc.edu.ar
\end{flushleft}

\section*{Abstract}
A series of recent works studying a database of chronologically sorted chess games 
--containing 1.4 million games played by humans between 1998 and 2007-- have shown 
that the popularity distribution of chess game-lines follows a Zipf's law, and that 
time series inferred from the sequences of those game-lines exhibit long-range memory 
effects. The presence of Zipf's law together with long-range memory effects was observed 
in several systems, however, the simultaneous emergence of these two phenomena were always 
studied separately up to now. 
In this work, by making use of a variant of the Yule-Simon preferential growth model, introduced 
by Cattuto et al., we provide an explanation for the simultaneous emergence of Zipf's law 
and long-range correlations memory effects in a chess database. We find that Cattuto's Model (CM) 
is able to reproduce both, Zipf's law and the long-range correlations, including size-dependent 
scaling of the Hurst exponent for the corresponding time series. CM allows an explanation for the 
simultaneous emergence of these two phenomena via a preferential growth dynamics, including a memory 
kernel, in the popularity distribution of chess game-lines. This mechanism results in an aging process 
in the chess game-line choice as the database grows. Moreover, we find burstiness in the activity 
of subsets of the most active players, although the aggregated activity of the pool of players 
displays inter-event times without burstiness. We show that CM is not able to produce time series 
with bursty behavior providing evidence that burstiness is not required for the explanation of the 
long-range correlation effects in the chess database.
Our results provide further evidence favoring the hypothesis that long-range correlations effects 
are a consequence of the aging of game-lines and not burstiness, and shed light on the mechanism that operates in the 
simultaneous emergence of Zipf's law and long-range correlations in a community of chess players.


\section*{Introduction} 
\label{intro}

In recent years the scope of statistical physics  
has been extended to other fields to study, for instance, how humans behave 
individually or collectively~\cite{watts2007twenty, lazer2009life, castellano2009statistical}.
In particular, the struggle of humans when playing games 
with a well defined set of rules provides a convenient experimental setup to understand human behavior and decision making 
processes~\cite{Prost12,Blasius09PRL,Ribeiro13PO, Sigman10FN,Chassy11PO,Sire09EPJB,
Guerra12PA,Petersen08EPL,Bittner09EPJB,BenNaim07EPL,Heuer09EPJB,Ribeiro12PRE,Xu15EPL}.
This is particularly convenient from the physics' point of view because, under the rules 
of a game, the variables of the behavior under study are highly constrained. 
Moreover, recent studies on game records~\cite{Sire09EPJB, Guerra12PA, Petersen08EPL, Bittner09EPJB, BenNaim07EPL, Heuer09EPJB, Ribeiro12PRE}
have shown that useful parallels can be established between the statistical patterns of game-based human behavior 
and well defined theories of physical processes.

The game of chess, which is viewed as a symbol of intellectual prowess,
is  particularly interesting~\cite{Blasius09PRL,Perotti13EPL,Schaigorodsky14PA}.
There are very large world-wide communities of chess players producing extensive game records providing a source of data 
suitable for large-scale analyses. 
Exploring chess databases, Blasius and T\"{o}njes \cite{Blasius09PRL} observed that the 
pooled distribution of chess opening-line weights follows a Zipf law with universal exponent. 
They explained these findings in terms of an analytical treatment of a multiplicative process. 
Moreover, in a previous work, where we studied the dynamics of the growth of the game tree, we have found 
that the emerging Zipf and Heaps laws can be explained in terms of nested Yule-Simon preferential growth 
processes~\cite{Perotti13EPL}. 

The study of long-range correlations and burstiness in systems where Zipf' s law is present 
is also of great research interest~\cite{Montemurro02F, Altmann09PO}. 
In recent studies, we found that temporal series generated from an empirical and chronologically ordered
chess database does exhibit long-range memory effects \cite{Schaigorodsky14PA}, 
establishing a striking similarity between chess databases and literary corpora\cite{Montemurro02F}.
These memory effects cannot be explained in terms of the multiplicative process, nor in terms 
of the nested Yule-Simon processes. In this sense, the mechanism for the game aggregation in the 
database needs new ingredients to  reproduce the observed long-range correlations. 

In the present work, we investigate how memory effects emerge in a ``corpus'' of chess games.
We tackle this problem following a result of Cattuto et al.~\cite{Cattuto06EPL}, who 
introduced a modification of the Yule-Simon process by incorporating a probabilistic 
memory kernel. 
Cattuto's model (CM) introduces memory  while preserving the long-tailed frequency distribution 
characteristic of Zipf's law~\cite{Montemurro02F}. However, the nature of the memory effects introduced 
by the kernel has not been yet explored. Here we show that CM reproduces the statistical properties 
observed in the chess database since the model does not only exhibit memory, but also long-range correlations 
and size effects. Moreover, we show that a bursty behavior is associated to individual players, and the most 
active set of players, but disappears when the full pool of players is analyzed, while long-range correlations 
prove to be more robust and are found in every analyzed set of games.

This work is divided in three sections. In Section \ref{sec:theory} we provide relevant 
information of the game of chess and the database. We also introduce Cattuto's model, the method for 
the analysis of long-range correlations and inter-event time distributions. 
In Section \ref{sec:result} we show and discuss the analysis of the data obtained from both, 
the generated and real database. Finally, in Section \ref{sec:discussion} we 
discuss the contributions of this work.

\section{Theoretical background}
\label{sec:theory}
\subsection{Zipf's law in chess}

A chess game can usually be divided in three stages, opening, middle-game and endgame. There are many 
specific openings ---move sequences in the opening stage--- that are very well documented because they are considered 
competitively good plays. As a consequence, in a chess database many of the recorded games have their first moves in common.
The openings evolve continuously and the complexity
of the positions determines their extension. 
The knowledge of opening-lines  is part of the theoretical
background of the chess players and how many opening-lines the players know in depth is certainly related to 
the expertise of the player. In practice, the extension of the opening stage cannot be precisely defined. 
In this work we will talk of opening-lines and game-lines to refer to sequences with the same number of moves.

\begin{figure}
\begin{center}
\includegraphics[scale=0.5]{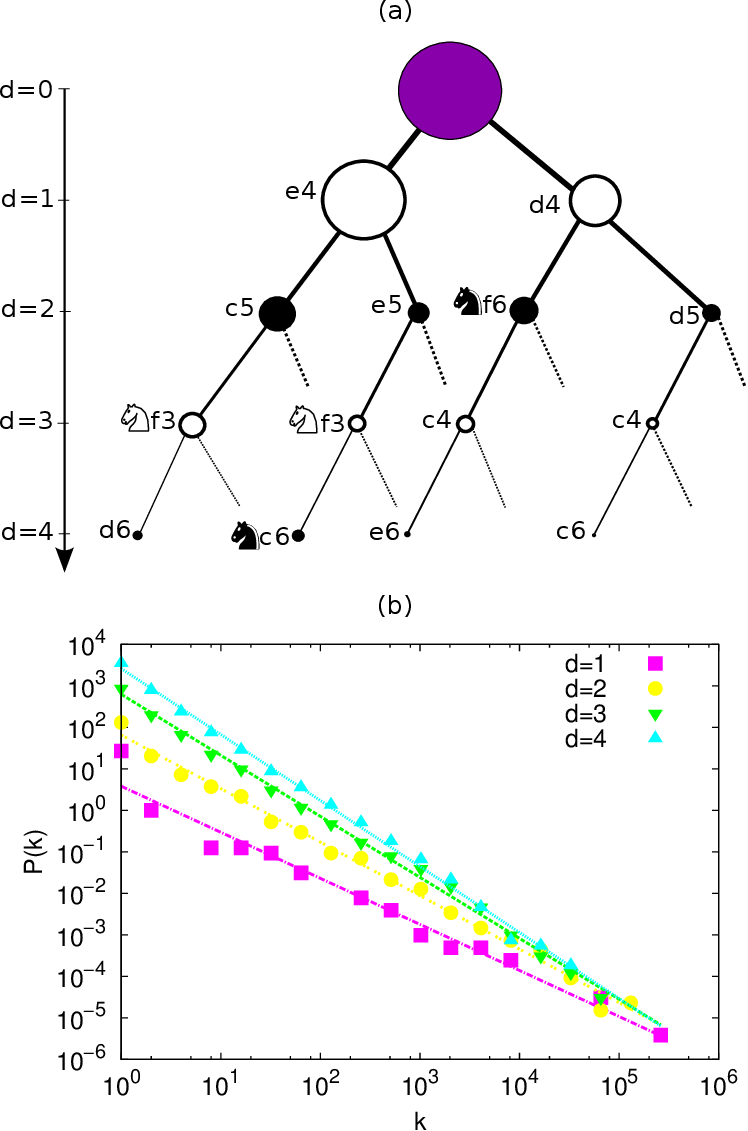}
\caption{\label{fig1} (a) Chess tree corresponding to the main opening-lines up to depth
$d=4$. The size of the nodes is proportional to their popularity. Here only the main lines are shown. 
(b) Distribution of popularities of the nodes at depth $d=1,2,3$ and $4$; these distributions are well fitted
by power laws $P(k) \propto k^{-\alpha}$ with $\alpha = 1.10 \pm 0.05,1.29 \pm 0.03,1.47 \pm 0.02$ and $1.59 \pm 0.02$
 ($R^2= 0.972, 0.993, 0.996$ and $0.997$), respectively.
Errors estimated by the fitting.}
\end{center}
\end{figure}

In the game of chess each possible move sequence, or game-line, can be mapped as one directed path in a corresponding game-tree 
(see Fig. \ref{fig1} (a)), where the root node is the initial position of the chess pieces in the board.
In the game tree each move is represented by an edge, and there is a one-to-one 
correspondence between game-lines and nodes.  The topological distance between the root and a node 
is the depth $d$ of the corresponding game-line.

Let us introduce some mathematical notation. A node, or game-line in the tree, is denoted by $g$. The popularity 
of a game-line $g$ ---i.e. the number of times $g$ appears in the database--- is denoted by $k_g$. In Fig. 
\ref{fig1} (a) we show a partial game-tree where the {\it popularity} is represented by the size of the vertices. 
This tree was computed from {\it ChessDB}\cite{foot1}, 
which contains around 1.4 million chess games played between the years 1998 and 2007.
This is the database we use for the rest of the analysis. The number of branches coming out of a node $g$ 
is denoted by $b_g$, and the depth of $g$ by $d_g$. The number of nodes at depth $d$ is denoted by $n_d$, and corresponds 
to the number of different game-lines that can be found in the database at depth $d$. Similarly, $N_d$ is 
the total number of games of the database that have reached depth $d_g=d$.

An average branching factor, or branching ratio, can be computed at each depth 
$d$ by using the formula
\begin{equation}
\avrg{b_d} = \frac{1}{n_d} \sum_{g:d_g=d}b_g = \frac{n_{d+1}}{n_d},
\end{equation}
where the summation goes over all existing nodes $g$ at depth $d$.
In practice, the chess database is continuously growing, i.e.
new games are incorporated to the database as time evolves.
Therefore, all these quantities change with time.
For practical reasons, we do not use the real time, but an ordinal time denoted by $t$.
In this sense, $g(t)$ is the game-line associated to the $t$-th game appearing in the database.
Similarly, $k_g(t)$ is 
the number of those $t$ games that have reached node $g$, $N_d(t)$ is the number of games that 
reached depth $d$ and $n_d(t)$ is the number of different game-lines among those $N_d(t)$ games~\cite{Perotti13EPL}.

From the statistical point of view the popularity of a given game-line depends on the 
 number of moves considered, i.e. the depth $d$ of the game. 
Blasius and T\"onjes found that the distribution of popularities follows a power law with an exponent that depends on $d$.
This means that there are few opening-lines which are very popular, and the rest are 
rarely played. We reproduce these results in Fig. \ref{fig1} (b), where the popularity 
distribution is shown for $d=1,2,3$ and $4$, and the curves are fitted using least square linear regression.
Clearly, the exponent increases with $d$, as it was reported~\cite{Blasius09PRL}.
A specific sequence of moves at a certain depth can be thought as a word, a string
in algebraic notation, and then the database as a literary corpus where the $t$-th game would correspond to the $t$-th word. 
In this way, analyzing the database at different depths is analogous to analyze different texts, all extracted 
from the same database, and all with different Zipf's exponents.

\begin{figure}
\begin{center}
\includegraphics[scale=0.6]{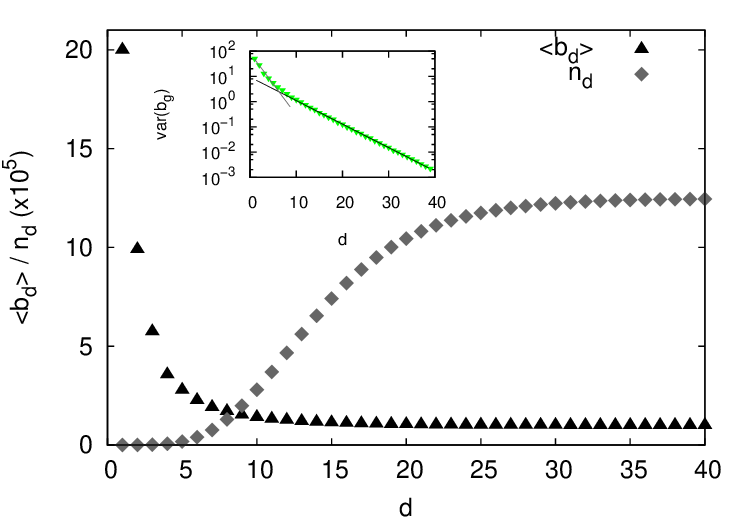}
\caption{\label{fig2} Average branching ratio, $\langle b_d \rangle$ and number of different game-lines 
$n_d$  as function of the depth level $d$ in the database. Inset: variance of the distribution of branches 
per node $b_g$ as function of the depth $d$ and linear fit of the two exponential regimes.
}
\end{center}
\end{figure}

The structure of the game-tree also depends on $d$.
In Fig.~\ref{fig2} the mean branching ratio is shown as a function of $d$.
The branching ratio quantifies both, the complexity of the game and the memory of 
the chess players when following the opening-lines. The branching 
ratio $\avrg{b_d}$ reaches a value $\approx 1$ for $d=25$, this means that the generation of new branches 
is negligible from this depth on, marking the beginning of the stage known as middle-game.
In Fig. \ref{fig2} we also show the number of different game-lines $n_d$ as function 
of the depth $d$.
At the beginning of the games, e.g. up to $d=4$, the number of game-lines followed 
by the players is relatively small, and a significant number of the players follow the 
most popular game-line. The statistical complexity of the game is reflected by the 
branching ratio $\avrg{b_d}$. Note that $\avrg{b_d}$ depends on the size of the database since new 
branches are generated as the database grows, and at the same time 
the popularity range depends on the depth $d$. 
Then, at $d=4$ we can capture both, the memory and the complexity of the game, since the
 more important opening-lines can be identified at this depth and the branching ratio is 
still higher than one ($\avrg{b_d} \approx 3.5$). Also, at this depth the exponent of the 
popularity distribution is $\alpha <2$ and then the range in which the popularity spans is more extensive 
than for higher depths.
Computing the distribution of the number of branches generated by each node $b_g$ for different values 
of the depth $d$ we have found that for lower 
depths ($d\leq 19$) the distribution is exponential, while for depths beyond $d=20$ a power law provides 
a better fit. However, it should be noticed that the range of fit covers around one order of magnitude 
only and, as a consequence, the power-law fit is not accurate.
In Fig.~\ref{fig2} (Inset) we show 
the variance of $b_g$ as a function of the depth. The fluctuations decay exponentially as $d$ increases.
Two regimes can be identified and the transition between them is related to the change of regime seen in  
$\avrg{b_d}$ and $n_d$.
Therefore, our analysis  will be restricted to the $6279$ opening-lines of length $d=4$ 
 found in the database.
In particular, we pay special attention to the most popular opening-line at this depth -- 
which is: \newgame \noindent \mainline{1. e4 e5 2.Nf3 Nc6} -- as it
represents nearly the $7.8 \%$ of the games in the database.
The reason for this is that several popular openings have these four initial moves in common.
For example: \variation{3. Bb5} (Ruy Lopez, by far the most popular), 
\variation{3. Bc4 } (Giuoco piano),
\variation{3. d4 } (Scotch opening) just to cite a few of them.
\subsection{Zipf's law models}
\label{models:section}
One of the first models able to explain the emergence of Zipf's law was introduced by Yule \cite{Yule25PTB}, which 
was devised to explain the emergence of power laws in the distribution of sizes of biological genera.
Later on, Simon \cite{Simon55B} introduced a similar, but less general, variation of the model~\cite{simkin2011re}, 
which fits more naturally in the context of Zipf's law. 
It is known as the Yule-Simon Model (YSM), and different 
variations re-emerged in the literature several times. The most recent variant, known as preferential 
attachment, became one of the most important ideas at the beginning of the development of complex networks theory~\cite{Barabasi99S}.
Cattuto et al. \cite{Cattuto06EPL} introduced another variant of the YSM, which includes memory effects by incorporating 
a probabilistic kernel, while at the same time preserving the long-tailed frequency distribution exhibited 
by the original YSM. 
The YSM applied to chess game-line generation is as follows. We begin with an initial state of $n_0$ game-lines, 
strictly speaking opening-lines at depth $d$. 
At each time step $t$ there are two options: i) to introduce a new game-line with probability $p$ or ii) to copy 
an already existing game-line with probability $\bar{p}=1-p$. 
In the latter case we have to determine which of the previous game-lines is to be copied. Note that since at each time 
step an opening-line is added, at time $t$ the total number of elements in the constructed database is $N=t+n_0$.
The probability of choosing a particular game-line, or opening-line, that has already occurred $k$ times is assumed to be
$\bar{p}k \pi(k,t)$, where $\pi(k,t)$ is the fraction of game-lines with popularity $k$ at time $t$. To fix ideas, lets take 
$N=5 \times 10^5$ and $100$ different game-lines with popularity $k$ at a time $t$, then $\pi(k,t)=\frac{100}{5 \times 10^5}=0.0002$. 
This means that, in the YSM, copying a certain game-line does not depend on how far back in time the game-line took place, 
but  only on how popular is the corresponding game-line up to the present time $t$.
For this reason, the process does not exhibit long-range memory effects. 
On the contrary, in CM, the probability of copying a previous game-line depends on how far back in time 
it occurred for the last time, taking into account the age of the game-line.
If the game-line occurred at time $t-\Delta t$, the probability is given by
\begin{equation}
\label{eq1}
Q(t,\Delta t)=\frac{C(t)}{\tau_c +\Delta t}.
\end{equation} 
In Eq.~\eqref{eq1}, $\tau_c$ is a time scale in which recently added game-lines have comparable associated probabilities, 
and it can be considered as a measure of the memory kernel extension. 
$C(t)$ is a logarithmic normalization factor. The probability 
distribution density for the popularity of the game-lines that results from this process is \cite{Cattuto06EPL}:
\begin{equation}
\label{eq2}
P_{CM}(k)=\frac{p}{(n_0 +pt)(K\alpha)k} \left[ \frac{\ln(A/k)}{K} \right]^{\frac{1}{\alpha }-1},
\end{equation}
where $\alpha=\bar{p}$, $K=\frac{1-\alpha}{\alpha \Omega}$, $A=e^{K t^{\alpha}}$ and $\Omega$ is a fit parameter.
Note that, strictly speaking, the mentioned models do not produce different sequences of moves, but elements that 
constitute an artificial database with the same distribution as the database. These models are of not use when trying 
to reconstruct the game tree, but are used to reproduce the statistical  properties of the system.

\subsection{Time series and correlations}
\label{cor:section}
In order to study the long-range correlations of the chronologically ordered set of games in the database, 
we map the set of game-lines of length $d=4$ to a discrete time series.

The particular assignation rule that maps the sequence of the $1.4\times10^6$ games of the database 
to a time series, can have a direct effect in the 
degree of persistence observed in the series \cite{Amit94F}. 
Specifically, long-range correlations are affected by both, the intrinsic properties of the database and the mapping code.
Therefore, we choose to work with different assignation rules in order to provide robustness to the results.
One of these rules, which is introduced in the analysis of literary corpora \cite{Montemurro02F} and was 
already employed in a chess database \cite{Schaigorodsky14PA}, is the Popularity Assignation Rule (PAR).
In PAR, each element $X(t)$ of the time series corresponds to the popularity at depth $d$ of the $t$-th 
game-line in the database over the entire record. 
In this work we introduce two more assignation rules for the analysis:
the Gaussian Assignation Rule (GAR), and the Uniform Assignation Rule (UAR).
GAR and UAR are random assignation rules, where a random number $X_g$ taken from the probability 
distribution function, Gaussian for GAR and uniform for UAR, is assigned to each game-line $g$ in the 
database.
In this way, the time series is $X(t)=X_{g(t)}$.
These random assignation rules are not expected to introduce spurious correlations.
Additionally, they have the advantage over PAR that the fluctuations in the values of the 
time series are bounded; large fluctuations in the values of a time series may induce spurious 
long-range memory effects~\cite{barunik2010hurst}.

There exists a wide variety of techniques used to detect long-range correlations in time series. 
However, not all of them are suitable to analyze all kinds of series, especially if they are 
non-stationary or exhibit underlying trends. 
Peng et. al.\cite{Peng92N} introduced the Detrended Fluctuation Analysis (DFA), a useful technique 
to detect long-range correlations in time series with non-stationarities. 
In the DFA method, a cumulated series $Y(i)=\sum_{t=1}^{i} X(t)$ is segmented into intervals of size $\ell$. 
Each segment $s$ of the cumulated series is fitted to a polynomial $Y^{(s)}_n(i)$ of degree $n$, 
and the fluctuation function is obtained with
\begin{equation}
\label{eq3}
F(\ell)
=
\sqrt{
\frac{1}{Z}
\sum_{t=1}^{Z} 
[Y(i)-Y^{(s_i)}_{n}(i)]^{2}
}.
\end{equation}
\noindent Here, $Z$ is the total number of data points in the time series, and $s_i$ is the segment 
of the $i$-th data point. A log-log plot of $F(\ell)$ is expected to be linear. 
If the slope is less than unity, it corresponds to the Hurst exponent ($H$). When $H = 0.5$ the cumulated time 
series, $Y(i)$, resembles a memoryless random walker. On the other hand, for $H > 0.5$ ($H < 0.5$), it 
resembles a random walker with persistent (anti-persistent) long-range correlations or memory effects.

\subsection{Inter-event time analysis}
\label{ss:theory:tau}

Inter-event time analysis is common to many natural system comprising, earthquakes \cite{Bak-PRL-02,Corral-PRL-2004},
sunspots \cite{Wheatland98AJ}, neuronal activity \cite{Keat01N}, and human behavior in 
general~\cite{barabasi2005origin,jo2015correlated}.
In particular, the time distribution of the opening-lines in a chess database can be analyzed in a similar manner 
than the occurrence of words in a text \cite{Altmann09PO}. 
All the game-lines up to depth $d$ can be enumerated according to its order of appearance in the chronologically ordered database. 
Specifically, we denote by $t_d \in \{1,2,...,N_d\}$ the sequence of ordinal times of appearance of the different 
opening-lines of length $d$.
Therefore, the  $j$-th inter-event time of an opening-line $g$ is defined as
\begin{equation}
\tau_j^{(g)} = t_d^{(g)}(j+1) - t_d^{(g)}(j),
\end{equation}
where $t_d^{(g)}(j)$ represents the time of the $j$-th appearance of the opening-line $g$.
If the opening-line $g$ occurs with frequency $\nu_{g} = N_d^{(g)} /N_d$, we can estimate the average 
inter-event time as $\avrg{\tau^{(g)}}\approx 1/\nu_{g}$.
Here, $N_d^{(g)}$ is the number of times the particular opening-line $g$ of length $d$ occurs in the database.
The mean inter-event time $\langle \tau^{(g)} \rangle$
is usually called the Zipf's wavelength in text analysis~\cite{Altmann09PO}, where $g$ represents 
a particular word.

The simplest point process for the analysis of inter-event times is the Poisson process.
In the context of chess game-lines it can be described as follows.
A particular opening-line $g$ occurs with a probability per unit of time equal 
to $\mu_g$, which we assume to be a constant. As a consequence, the inter-event time distribution 
for the opening-line $g$ is the exponential distribution $f(\tau^{(g)}) = \mu_{g}\exp(-\mu_{g}\tau^{(g)})$.
Here, the rate $\mu_{g}\approx \nu_{g}$.
The relation is approximate because of two reasons.
Firstly, Poisson processes are defined for a continuous time, while for the chess database we are considering a discrete time.
Secondly, the fraction $\nu_{g}$ corresponds to a finite number of events, while a Poisson process 
describes an infinitely large stationary process.
Besides this, the approximation $\mu_{g}\approx \nu_{g}$ works well as long as $\nu_{g}\ll 1$ and $N_d \gg N_d^{(g)} \gg 1$.

Let us simplify the notation by writing $\tau$ instead of ${\tau^{(g)}}$, when there is no 
need to speak about a particular game-line $g$.
In the empirical analysis of the data, it is convenient to use the complementary cumulative probability 
density $F(\tau) = \int_\tau^\infty f(\tau')d\tau'$ instead of a direct application of the probability 
density $f(\tau)$. This is for practical reasons; the function $F(\tau)$ is usually simpler than $f(\tau)$.
For example, a deviation of $F(\tau)$ from an exponential behavior indicates the presence of memory-effects.
In the case of words in a text, this deviation 
is usually well described by the single parameter stretched exponential distribution, or Weibull function \cite{Altmann09PO}, 

\begin{equation}
\label{eq5}
f(\tau)= \frac{\beta}{\tau_0} \left(\frac{\tau}{\tau_0}\right)^{\beta-1} e^{-(\frac{\tau}{\tau_0})^\beta}.
\end{equation}

\noindent For this distribution $\langle \tau \rangle = \tau_0 \Gamma(\frac{\beta +1}{\beta})$,
where $\Gamma$ is the Gamma function and $0<\beta \leq 1$.
The corresponding cumulative distribution is,
\begin{equation}
\label{eq6}
F(\tau) = e^{-(\frac{\tau}{\tau_0})^\beta}. 
\end{equation}

\noindent If $\beta$ deviates from one, the presence of burstiness in the time series is implied.
A burst corresponds to an increase in the activity levels over a short period of time
followed by long periods of inactivity \cite{Goh08EPL}, and as the value of $\beta$ approaches zero
the appearance of bursts in the time series increases.
To test if the cumulative distribution of inter-event times follows or not a stretched exponential, 
it is useful to plot $-\log(F(\tau))$ as function of $\tau$ in a $\log-\log$ scale \cite{Bunde05PRL,Altmann09PO}.
In this plot a stretched exponential becomes a straight line where the slope is the burstiness 
exponent $\beta$.  

The deviation of $F(\tau)$ from a Poisson process can also be characterized with the coefficient of 
variation $\sigma_{\tau}/ \langle \tau \rangle$, where $\sigma_{\tau}$ is the standard deviation of the inter-event times. 
We use the coefficient of variation to compute the burstiness parameter $B$ as~\cite{Goh08EPL}, 
\begin{equation}
B = \frac{(\sigma_{\tau} / \langle \tau \rangle-1)}{(\sigma_{\tau} / \langle \tau \rangle+1)}= \frac{\sigma_{\tau}  - \langle \tau \rangle}{ \sigma_{\tau} + \langle \tau \rangle}.
\end{equation}
This parameter is greater than zero for a bursty dynamics and less than zero when dynamics becomes regular. 
When $B=0$ there is neither burstiness nor regularity.

\section{Results}
\label{sec:result}

\subsection{Fitting the parameters of the models}
\label{ss:param_fit}

Let us begin by fitting the model parameters in order to reproduce some basic statistical properties of the database.
The parameters to be fitted are: $p$ in the case of the YSM, and $p$ and $\tau_c$ for CM. Artificial databases 
of $N=10^6$ elements are generated by applying both models' update rules, introduced in Section~\ref{models:section}, $N$ times.

The appropriate value of the parameter $p$ can be directly estimated from the database by using the formula,
\begin{equation}
\label{eq4}
p 
\approx
\frac{n_d(t_{\mathrm{total}})}{t_{\mathrm{total}}}.
\end{equation} 
This estimation is only valid as a first approximation since we implicitly assume that 
$p$ is a constant function of $t$ but,
in fact, the number of different game-lines grows in time according to the Heaps' law~\cite{Perotti13EPL}, 
and not linearly as in Eq.~\eqref{eq4}. 
However, in order to keep the analysis simple, we choose to work within the approximation of constant $p$, 
as this is the case for YSM and CM.
For the case of $d=4$, the estimated value is $p=0.005$. It is worth mentioning that for larger values 
of $d$ the approximation of constant $p$ is not as appropriate~\cite{Perotti13EPL}.

To obtain an appropriate value for the parameter $\tau_c$ ---a parameter of CM only--- 
we vary $\tau_c$ until CM is able to reproduce the average inter-event time 
$\avrg{\tau^{(g^*)}}$
of the most popular game-line $g^*$  in the database at $d=4$.
Then, provided that $p$ is given by Eq.~\ref{eq4}, the best approach of CM occurs for
$\tau_c = 96$, and for this value CM gives $\avrg{\tau^{(g^*)}} = 12.41$.
Furthermore, the most popular game-line generated by CM, represents the $8.1\%$ of the game-lines; 
a value close to the empirical one which is $7.8\%$.

The YSM also provides a prediction for $\avrg{\tau^{(g^*)}}$.
However, when $p$ is given by Eq.~\ref{eq4}, the prediction is
$\avrg{\tau^{(g^*)}} = 7.68$; a value considerable smaller than the observed in the database.
The correct prediction can be obtained anyway, if we set $p=0.1$, which is a value considerable larger than the obtained from Eq.~\ref{eq4}.
In other words, the YSM is not able to simultaneously fit the empirical values of $p$ and $\avrg{\tau^{(g^*)}}$, while CM does.
This is expected, as CM has an extra fitting parameter.

For comparison, we summarize in Table~\ref{tab:1} the different values of $p$ and $\avrg{\tau^{(g^*)}}$ obtained 
from the database and the models.

\subsection{Comparing the models}

After setting the model parameters $p$ and $\tau_c$, we test the models against complementary statistical properties 
measured to the chess database, such as the popularity distribution and the presence of long-range memory effects.

\subsubsection{Popularity distribution}

\begin{figure}
\begin{center}
\includegraphics[scale=0.95]{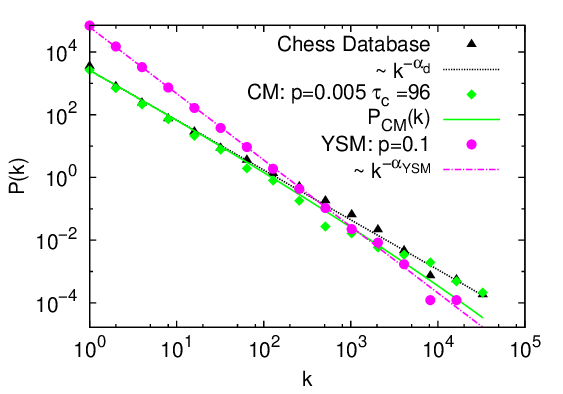}
\caption{\label{fig3} Log-log plot of the distribution of popularities: measured in the database 
(black triangles), fitted with $P(k) \propto k^{-\alpha}$ and exponent $\alpha_d=1.59 \pm 0.02$ ($R^2=0.997$) (dotted black line); 
generated with CM, $p=0.005$ and $\tau_c=96$ (green diamonds) fitted with $P_{CM}(k)$ 
(see Eq. \eqref{eq2}) with parameter $\Omega=1.5 \pm 0.3$ (full green line); and generated with YSM model, 
$p=0.1$ (magenta circles) and fitted with $P(k)$ and exponent $\alpha_{YSM}=2.12 \pm 0.03$ ($R^2=0.997$) (magenta point dashed line).} 
\end{center}
\end{figure}

In the following, the parameters of the models are fixed to the values obtained
in the previous section.
In Fig.~\ref{fig3} we show the distribution of popularities $P(k)$ of the
YSM, CM and the database (opening-lines of length $d=4$).
The YSM model produces a power-law distribution, with an exponent 
very close to $2$, which is expected in this process for small values of $p(=0.005)$~\cite{Simon55B}.
The distribution obtained from CM shows a gentle curvature, and is very well fitted 
by the theoretical expression of Eq.~\eqref{eq2}. The distribution $P(k)$ of the 
database is much closer to that obtained with CM than with YSM. 
A similar popularity distribution can be obtained with the YSM if we relax the restriction where $p$ is given by Eq.~\ref{eq4}.

\subsubsection{Hurst exponent}

In order to analyze the presence of long-range correlations, we measured the Hurst exponent ($H$) of 
time series derived from the models and the empirical data.
The time series are obtained using three different assignation rules: PAR, GAR and UAR, 
and the Hurst exponent is computed with a linear DFA method~(see Section~\ref{sec:theory}).
Again, for CM we set the parameters obtained in Section \ref{ss:param_fit}. It is worth 
mentioning that long-range correlations are present in time series constructed from the database 
for depths greater and lower than $d=4$~\cite{Schaigorodsky14PA}.

\begin{figure}
\begin{center}
\includegraphics[scale=0.95]{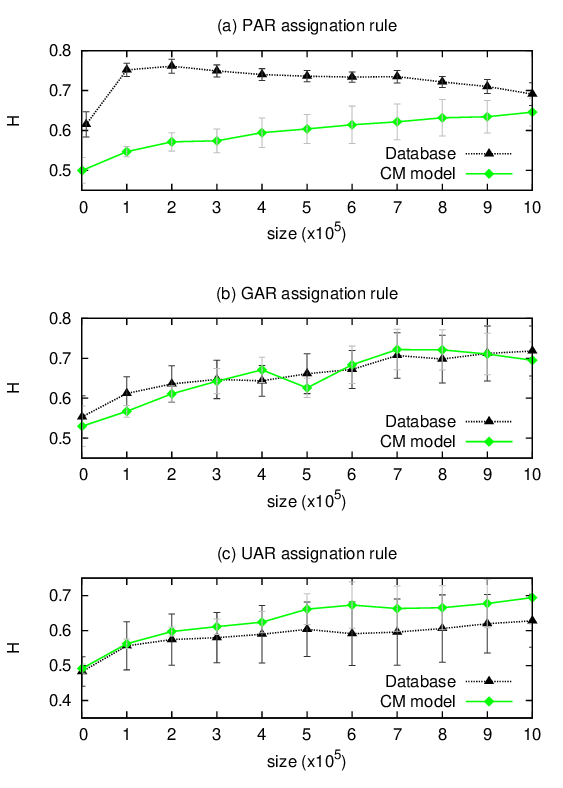}
\caption{\label{fig4} Hurst exponent obtained by the DFA method as a 
function of the length of the time series in the database (dotted black line with triangles), 
and generated with CM, $p=0.005$ and $\tau_c=96$ (full green line with diamonds): 
(a) using the PAR; (b) using the GAR; and (c) 
using the UAR.
}
\end{center}
\end{figure}

Consistently with a lack of long-range time correlations, the Hurst exponent corresponding to the 
YSM is close to $0.5$. Moreover, this result (not shown) is independent of both $p$ and
the assignation rule.
\\
In Fig. \ref{fig4} (a) we show the Hurst exponent as function of the length of the 
time series, using the PAR for CM and the database.
The time series generated with CM exhibits both, long-range correlations  
and size effects,
behaving similarly to the database. The value of $H$ grows up to $0.69$ for the database, 
and up to a similar value ($0.65$) in the case of CM. 
The tendency is different in both cases, while in the database the Hurst exponent becomes large for short 
time scales,  it grows steadily in CM.

As mentioned in Section \ref{cor:section}, large fluctuations in the values of the time series $X(t)$ might 
introduce spurious long-range memory effects, i.e. values of $H$ significantly different from 0.5. 
Since the popularity distribution is long-tailed 
---in both the model and the empirical database--- the PAR rule leads to large fluctuations in the values of $X(t)$.
In order to test the influence of these fluctuations we repeated the calculations 
of $H$ using time shuffled series $X_{\mathrm{shuff}}(t)$. 
The Hurst exponents obtained after the shuffling are very close to 0.5~\cite{Schaigorodsky14PA} (not shown), 
and thus, the large fluctuations in $X(t)$ are not the cause of the observed long-range correlations.
\\
In a similar manner, we can check if the condition $H>0.5$ 
persists when the fluctuations in the values of the time series $X(t)$ are bounded. 
For that purpose, we used the others assignation rules, UAR and GAR, as they lead to time series with finite  variance.
In Fig. \ref{fig4} we also show the Hurst exponent as a function of the size of 
the analyzed time series for this two more assignation rules; GAR Fig.~\ref{fig4} (b) 
and UAR Fig.~\ref{fig4} (c). 
The obtained Hurst exponents are $H>0.5$ almost everywhere.
Therefore, the emergence of long-range correlations is robust against the
choice of the assignation rules.
In particular, we obtained a nice agreement between the database and CM
model when using the GAR. Also, it is worth mentioning that it has been found empirically 
that DFA tends to be more robust for Gaussian processes~\cite{Bryce12SR}.

The error bars in Fig.~\ref{fig4} 
(a) for the database are the errors resulting from the linear fitting of $F(l)$, while for CM
we have computed 10 realizations of the model, and the error bars reflect the dispersion of the
calculated values of $H$.  
However, in panels (b) and (c), which correspond to the GAR and UAR, 
we estimated the errors using 10 different random assignations for each UAR and GAR,
for both CM and the database.
The different assignation rules lead to different values of the 
Hurst exponent; both, in the model and in the database.
This implies that the existence of long-range correlations is a robust feature from a qualitative 
point of view but, the Hurst exponent is not a quantity independent of the assignation rule used 
to produce the time series.

\subsection{Burstiness}

In this subsection, we study another aspect of the existence of non-trivial memory effects, 
regarding the occurrence of specific opening-lines in the chess database.
Namely, we analyze the existence of burstiness by following ideas used in the analysis of texts~\cite{Altmann09PO}.

Before continuing, we remark a couple of issues.
First, our calculations assume that the games in the database are uniformly distributed across the time line.
Our results indicate this assumption is correct.
Second, although the games in the chess database are chronologically ordered, the time resolution of the database 
is one day, with an average of about $\approx 3600$ games per day which are randomly shuffled. For this reason, 
we expect the sequence of inter-event times $\tau$ to be uncorrelated at the short scale of $t$.

\subsubsection{Complete database}

We begin the study of burstiness by analyzing the activity of the most popular opening-line, involving the whole pool of players.
We study both the database and the models.

\begin{figure}
\begin{center}
\includegraphics[scale=0.65]{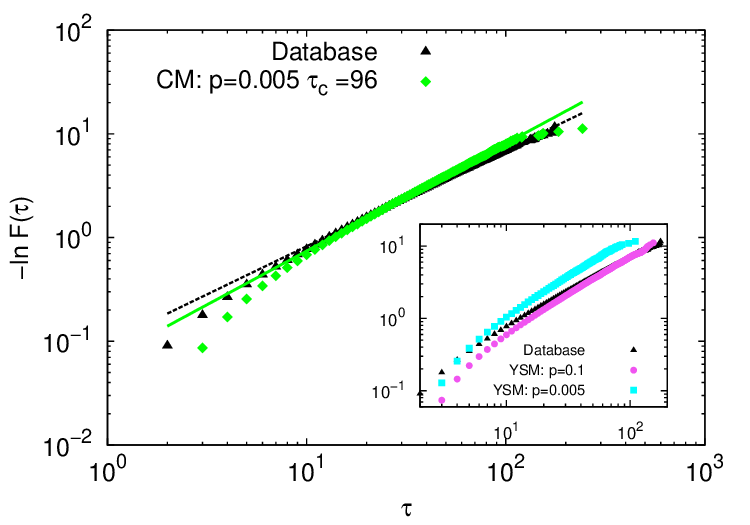}
\caption{\label{fig5} Cumulative distribution of inter-event times of the most popular 
opening-line measured in the database (black triangles) and generated with CM for $p=0.005$ and $\tau_c =96$ 
(green diamonds). The lines are fits according to Eq.~\eqref{eq6} with $\beta=0.927 \pm 0.003$ ($R^2=0.999$) for the database (black dashed line) 
and for CM with $\beta=1.035 \pm 0.003$ ($R^2=0.999$) (green full line). Inset: cumulative frequency distribution of inter-event times measured 
in the database (black triangles), YSM for $p=0.1$ (magenta circles) and $p=0.005$ (cyan squares). }
\end{center}
\end{figure}

We analyze the cumulated inter-event times distribution $F(\tau)$~(see~\ref{ss:theory:tau}) 
corresponding to the most popular opening-line at depth $d=4$.
In Fig. \ref{fig5}, we plot $-\ln F(\tau)$ as function of $\tau$ in a $\log-\log$ scale for the case of the empirical database.
Analogous plots for CM and YSM (Inset) are also shown.
All cases display a linear regime, admitting a good fit of the Weibull distribution of Eq.~\ref{eq6}.
The exponent of the Weibull and the burstiness parameter results  $\beta \approx 1$  and $B \approx 10^{-2}$, 
respectively, in all cases, indicating the absence of burstiness. 
The linear regime was estimated by eliminating points from both ends of the curve $-\ln F(\tau)$ and doing a linear 
regression for each subset, until the coefficient of determination $R^2$ reached a stable value with a tolerance 
of $10^{-4}$. 

Because there is no burstiness, the other parameter of the Weibull distribution, $\tau_0$, should fit to the characteristic 
time $\tau_P$ of an associated Poisson process.
We corroborated this by computing a simple approximation, derived for the YSM, but which also works for CM.
In the YSM model, the probability $P_t$ that an already existing game-line $g$ is repeated between times $t$ and $t+\delta t$, 
is given by $P_t \approx ((1-p)N^{(g)}_d(t)/N_d(t))\delta t$, where $N^{(g)}_d(t)$ is number of game-lines $g$ up to depth $d$ that have 
been played until time $t$ (see \ref{ss:theory:tau}).
After a transient time, $N^{(g)}_d/N_d$ is expected to be nearly constant ---to be precise, a slowly varying quantity---.
Therefore, $P_t = \delta t / \tau_P$, where 
$\mu := 1/\tau_P \approx (1-p)N^{(g)}_d/N_d$ is the event rate of the mentioned Poisson process,
and
\begin{equation}
\label{eq7}
\tau_P \approx \frac{ N_d(t_{total})} {N_d^{(g)}(t_{total}) (1-p) }
\end{equation} 
the corresponding characteristic time.
Here, $t_{total}$ is the number of games in the whole database.
As summarized in Table~\ref{tab:1}, it is clear that there is a good agreement between the fit $\tau_0$ and the estimation $\tau_P$.

\begin{table}
\caption{
Summary of the results corresponding to the inter-event time distributions of Fig.~\ref{fig5} 
and Fig.~\ref{fig6}. 
}
\label{tab:1}
\begin{center}
\begin{tabular}{l|c|c|c|c}
Data    & $p$  & $\beta$ & $\tau_0$& $\avrg{\tau^{(g^*)}} \approx \tau_P$ \\
\hline
\hline
Database  & 0.005 & $0.927 \pm 0.003$  &   $13.0 \pm 0.5$   & 12.82 \\
CM     & 0.005 & $1.036 \pm 0.002$ &    $12.9 \pm 0.8$    & 12.41 \\
YSM    & 0.1  & $1.031 \pm 0.003$ &   $15.4 \pm 0.6$     & 14.12 \\
YSM    & 0.005 & $1.059 \pm 0.005$ &   $8.2 \pm 0.6$    &  7.68 \\
Player & 0.09 &  $0.583 \pm 0.004$ &   $4297 \pm 200$  & 89.75 \\
\end{tabular}
\end{center}
\end{table}

\subsubsection{Single player analysis}

According to the results of previous sections, the inter-event times of the most popular game-line indicate a slight, 
or absence, of burstiness when the whole pool of players is considered. 
In order to shed light on this point,
we analyze if there is burstiness in the activity of the game-lines of a single player. 

\begin{figure}
\begin{center}
\includegraphics[scale=0.6]{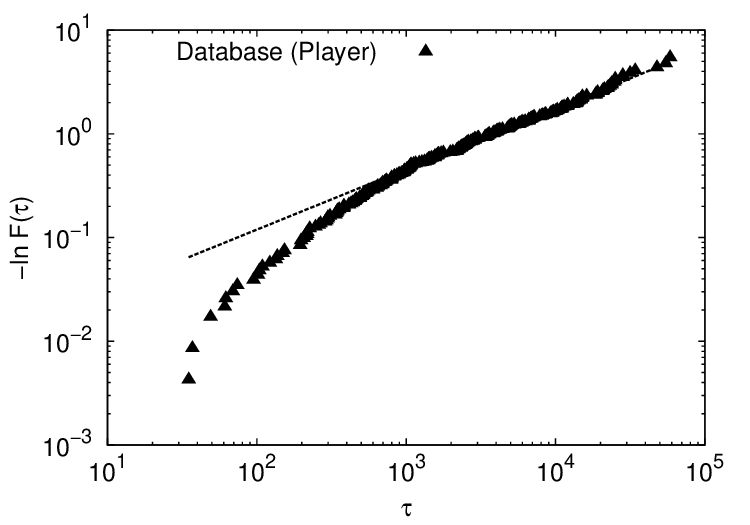}
\caption{\label{fig6} Cumulated inter-event time distribution for a single player 
(black triangles). Also, corresponding fits according 
to Eq.~\eqref{eq6} with $\beta=0.583 \pm 0.004$ ($R^2=0.993$) for the player (black dashed line).  
}
\end{center}
\end{figure}

For this analysis, we choose the most active player in the database, who has played $1377$ games. 
We keep the original time indices.
The measured value of $p$ for this player is equal to $0.09$.
In Fig. \ref{fig6} we show the cumulated inter-event time distribution for the most popular opening-line according to this player. 
The fitted slope of the linear regime is $\beta=0.583 \pm 0.004$ and the burstiness parameter  is $B=0.22$, indicating the 
presence of a bursty activity. 
Also, the values obtained for $\tau_0$ from the fitting of Eq.~\eqref{eq6} and $\tau_P$ from Eq.~\eqref{eq7}
are very different (see Table~\ref{tab:1}).
As a consequence, a Poisson process is not a good approximation for the inter-event time distribution of the single player. 

As we already showed, CM does not generate burstiness, hence we cannot reproduce the results 
of a single player with this model.

\subsubsection{Data aggregation}

\begin{figure}
\begin{center}
\includegraphics[scale=0.7]{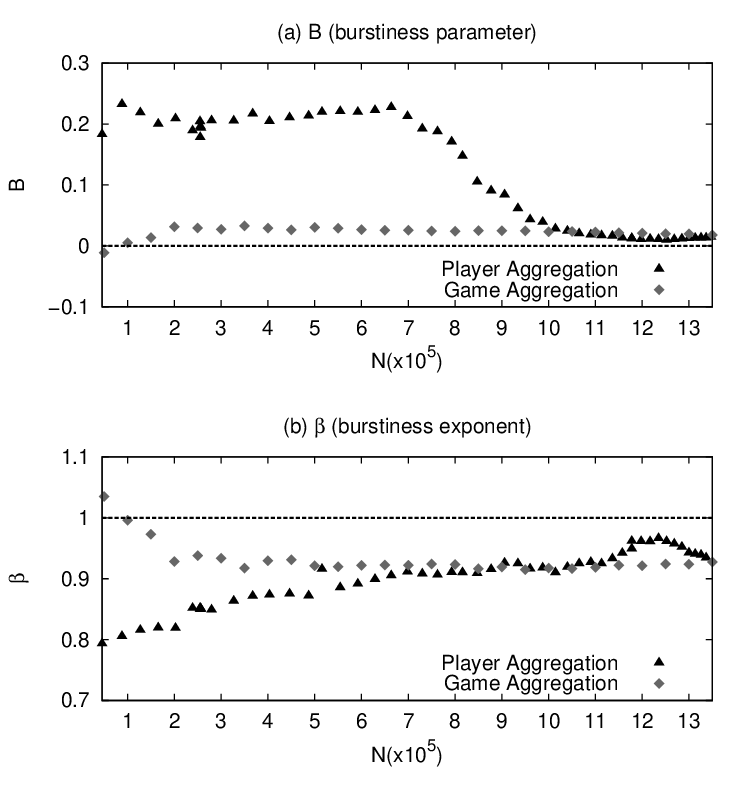}
\caption{\label{fig7} (a) Burstiness parameter $B$ and (b) burstiness exponent $\beta$  
as a function of the length of the time series for the database growing by Player Aggregation (black triangles) and  by chronological 
game aggregation (gray diamonds). The error bars are not shown because they are of the same size as the points.
}
\end{center}
\end{figure}

We also studied the bursty behavior during data aggregation. 
We grew the database in two ways, by player aggregation, and by chronological game aggregation.
In the first case, we ordered the players according to the number of games they have in the database; from most active to least active. 
We grew the database adding groups of players, each group representing $5 \times 10^4$ games, until we completed the full database. 
In this way, the most active players are included first during the growing process. 
In the second case, game-lines were added chronologically in sets of size $5 \times 10^4$ games, also until the whole database is completed.
For these two data aggregation processes we calculated the burstiness parameter $B$ and the burstiness exponent $\beta$. The resulting 
values as function of the size of the database are shown in Fig.~\ref{fig7}. 
In this figure, it is noticeable that the points in the players aggregation are not uniformly distributed across the time scale.
This is because when we add new players, many of the games played by them are already in the growing database if the opponents 
involved in the games were already added. 
The effect is more significant when the growing 
database reaches a considerable size. 

The analysis of Fig. \ref{fig7} reveals that for both, player and game aggregation, $\beta$ and $B$, stabilize to the values of the complete database,
meaning that in both cases the burstiness disappears at some point. Notice, however, they reach the final value at different times.  
Under game aggregation, the values stabilize at approximately $2\times 10^{5}$ while under player aggregation, the values stabilize much later.
This is a new indication that the bursty dynamics in the whole database is related to the behavior of individual players. 
In fact, during the chronological game aggregation, the incorporation of a small number of game-lines (the first $5\times 10^4$) is already sufficient to include a relatively large fraction of players (around $12\%$ of the total).
This is the reason for which the bursty behavior disappears at a short time scale under game aggregation. 
In player aggregation, however, we have  
$57$ players in the first set, and the burstiness stabilizes when the number of players is $\approx 1000$, which corresponds to $ 10^6$ games. 
The remaining $4 \times 10^5$ game-lines correspond to players with few games --most of them with less than ten--.
These players cannot exhibit burstiness by themselves and therefore erase the bursty behavior in the database.  
Finally, we have checked that all players with an extensive record of games exhibit a bursty dynamics.   

\section{Discussion}
\label{sec:discussion}

In this work we analyzed a chess database that exhibit a power law popularity distribution of the opening 
lines and long-range correlations in the aggregation of new games during the growth process of the database. 
As a complementary test of the memory present in the database we also looked for the presence  
of a bursty dynamics in the apparition of the most popular opening-line. 

We have analyzed the database within the framework of two models based on a preferential grow mechanism, 
Yule-Simon Model (YSM) and Cattuto's Model (CM), 
since both models allow to generate artificial databases for opening-lines with power law distribution. 
In particular, the model proposed by Cattuto et al. includes a memory kernel to this process.
Our analysis demonstrates that both models can reproduce to a certain extent the popularity distribution 
obtained from the chess database, but CM seems to be more realistic since it reproduces the power law distribution 
of opening-lines  using  the  measured value for the probability of introducing  a new opening-line. 
In addition, due to the memory kernel, we show that CM is also able to reproduce the long-range correlations
and size effects observed in the database while, as expected, YSM lacks memory.  
Finally,  neither CM or YSM generate a bursty dynamics, for this reason  CM is
adequate to model the whole database but not to analyze this specific dynamic in individual players.   

Specifically, we showed that the popularity distribution of opening-lines at depth $d = 4$ is
well described by CM using the parameters $p$ measured in the database, 
and $\tau_c$ determined by the fitting of the mean inter-event time of the most popular 
opening-line. YSM reproduces this popularity  distribution but using a value of $p$ that is much greater
than the measured value. 
Furthermore,  CM exhibits long-range  correlation and size effects, independently of the assignation 
rule used to construct the time series, though the degree of persistence does depend on the assignation.
In particular,  CM reproduces the size effects and persistence observed in the database when the Gaussian Random Assignation
rule (GAR) is used.
This suggests that there are some underlying  correlations relating the popularity of chess game-lines
and the corresponding generated time series, which are not captured by CM. 

Furthermore,  we found from the fit of the cumulated inter-event time distribution of the complete chess database ---using the most 
popular opening-line--- that we cannot assure the presence of burstiness. 
The exponent of the Weibull distribution is essentially one ($\beta = 0.927$) and, therefore, the dynamics can be well 
approximated by a Poisson process. This is  validated further using the burstiness parameter since  $B \approx 10^{-2}$ 
and also because the fitted value of $\tau_0$  (see Eq. ~\eqref{eq6}) and the computed value of $\tau_P$ (see Eq.~\eqref{eq7}) 
are very similar. 
In addition, since the inter-event times of both CM and YSM are also
well described by a Poisson process, then the inter-event times of the database can be reproduced using 
both models. The lack of burstiness in CM suggests that burstiness phenomena --which is frequently used 
to explain the emergence of long-range memory effects-- may play a marginal role in the presence of long-range correlation 
effects. Instead, long-range correlations  emerge due to the aging of game-lines as database grows.  

Although the behavior of the complete set of players does not exhibit bursty behavior, 
the analysis of the inter-event times of individual players, provided that they have 
a sufficient long number of played games, does have bursty
behavior. This indicates that the lack of burstiness at the group level is a consequence of 
the aggregation of the data. The two growing mechanisms tested in this
work, player aggregation and game aggregation confirm this idea, since the first preserves
burstiness in a much longer time scale than the second. This may be due to underlying correlations between players, e.g. the 
two contenders in a game tend to have similar ELO rankings, and then the game-lines are necessarily correlated.

Concluding, the long-range memory observed in the opening-lines of the growing chess database can be 
well described with CM. However, new ingredients are necessary to explain the bursty dynamics
present in some growing stages of the database and in individual players. Moreover, further studies 
are necessary to test the implications of the memory kernel in the mechanism of the database generation.

\section*{Acknowledgments}

This work was partially supported by grants from CONICET (PIP 112 201101 00213), SeCyT--Universidad 
Nacional de C\'ordoba (Argentina)
J.I.P acknowledges support from: FET IP Project MULTIPLEX nr. 317532. FET Project SIMPOL nr. 610704, 
FET project DOLFINS nr. 640772. 

\nolinenumbers

%
%
%


%

\end{document}